\documentclass[12pt]{article}
\usepackage{epsf}
\usepackage{graphicx}
\usepackage[small]{caption2}
\usepackage[T1]{fontenc}
\usepackage{times}

\begin{document}
\title{
Simulations of metastable decay in 
two- and three-dimensional models with microscopic dynamics
}
\author{
M.A.\ Novotny$^{a,0}$, P.A.\ Rikvold$^{a,b,0}$,\\ 
M.~Kolesik$^{c}$, D.M.\ Townsley$^{a,b,1}$, R.A.\ Ramos$^d$\\
{\normalsize
$^a$Supercomputer Computations Research Institute,}\\ 
{\normalsize
Florida State University, Tallahassee, FL 32306-4130, USA}\\
{\normalsize
$^b$Center for Materials Research and Technology and 
Department of Physics,}\\
{\normalsize
Florida State University, Tallahassee, FL 32306-4350, USA}\\
{\normalsize
$^c$Department of Mathematics, University of Arizona,
Tucson, AZ 85721, USA}\\
{\normalsize
$^d$Department of Physics, University of Puerto Rico, 
Mayaguez, PR 00681, USA}
}
\date{\today}
\maketitle
\footnotetext[0]{
Corresponding authors. Fax: +1-850-644-0098; 
e-mails: novotny@scri.fsu.edu, rikvold@scri.fsu.edu 
} 
\footnotetext[1]{
Present address: 
Department of Physics, University of California, 
Santa Barbara, CA 93106-9530, USA.
}

\begin{abstract}

We present a brief analysis of the crossover 
phase diagram for the decay of a metastable phase in a simple 
dynamic lattice-gas model of a two-phase system.  We illustrate the 
nucleation-theoretical analysis with 
dynamic Monte Carlo simulations of a kinetic 
Ising lattice gas on square and cubic lattices.  
We predict several regimes in which the metastable lifetime has 
different functional forms, and provide estimates 
for the crossovers between the different regimes.  
In the multidroplet regime, the 
Kolmogorov-Johnson-Mehl-Avrami theory for the time 
dependence of the order-parameter decay and the 
two-point density correlation function allows extraction 
of both the order parameter in the metastable phase and 
the interfacial velocity from the simulation data.  
\end{abstract}

\noindent
1999 PACS numbers: 
64.60.My, 
64.60.Qb, 
81.30.-t, 
05.50.+q  
\clearpage

\section{Introduction}
\label{sec:INTRO}

The decay of a metastable phase through nucleation and growth of 
droplets of a more stable phase is a common feature of phase-transformation 
processes in both natural and technological settings. It occurs both 
in systems where the phases involved are non-crystalline 
and crystalline \cite{COMMENT1}.  
Often such processes are well described by a combination of nucleation 
theory \cite{ABRA74} 
and the Kolmogorov-Johnson-Mehl-Avrami (KJMA) theory 
\cite{KOLM37,JOHN39,AVRAMI}, which describes nucleation 
followed by irreversible growth of the nucleated droplets \cite{RIKV94}. 

Here we present some recent theoretical results on the effects 
of finite system size in simple model systems with microscopic dynamics, 
which undergo decay of a metastable phase through homogeneous nucleation and 
growth. These results are illustrated and confirmed by extensive dynamic 
Monte Carlo (MC) 
simulations. The simulations utilize several advanced algorithms 
\cite{BORT75,KOLE98A,KOLE98B,NOVO99B,NOVO95} 
that enable numerical study of extremely long-lived metastable phases 
{\it without\/} changing the 
underlying dynamics. Although the models that we study are highly simplified, 
there is evidence from magnetic systems that the predicted finite-size 
effects are experimentally observable \cite{RICH94,RIKV97B}. 
Moreover, our data-analysis methods should also be 
applicable to experiments. 

The rest of this paper is organized as follows. The theoretical model is 
presented in Sec.~\ref{sec:MOD}, and 
relevant aspects of the nucleation and KJMA theories are presented in 
Sec.~\ref{sec:NG}. Results of MC simulations are given in 
Sec.~\ref{sec:NR}. In particular, we present a 
``metastability phase diagram,''
metastable lifetimes for two- and three-dimensional systems, and 
measurements of the metastable order parameter and the propagation 
velocities of domain walls in two-dimensional systems. A discussion and 
conclusions are given in Sec.~\ref{sec:CONC}.

\section{Theoretical Model}
\label{sec:MOD}

We consider a simple lattice-gas model of a two-phase system, defined by 
the effective Hamiltonian 
\begin{equation}
{\cal H}_{\rm LG} = - \phi \sum_{\langle i,j \rangle} c_i c_j 
- \mu \sum_i c_i \;.
\label{eq:HLG}
\end{equation}
Here $c_i$$\in$$\{0,1\}$ is a local concentration variable at the $i$th 
site of a $d$-dimensional 
hypercubic lattice with periodic boundary conditions, 
$\phi$ is an attractive interaction between 
neighboring sites, and $\mu$ is a chemical potential favoring 
$c_i$$=$$1$. The sums $\sum_{\langle i,j \rangle}$ and $\sum_i$ run over 
all nearest-neighbor pairs and all lattice sites, respectively, 
and the lattice constant is taken as unity. 
The order parameter conjugate to $\mu$ is the dimensionless 
density (for $d=2$ often called ``coverage''), 
$\theta = {\cal N}^{-1} \sum_i c_i$,
where $\cal N$ is the total number of sites. Below a 
critical temperature $T_c$, 
two uniform phases coexist 
along a line of first-order phase transitions at 
$\mu = \mu_0 = - d \phi$. 
Here 
$\mu$ can be interpreted as an (electro)chemical potential, 
or related to an (osmotic) pressure $p$ as 
$\mu - \mu_0 = k_{\rm B} T \ln (p/p_0)$ where $k_{\rm B}$ is Boltzmann's 
constant, $T$ the absolute temperature, and $p_0$ the pressure at 
coexistence. Analogous relations can be made to 
supersaturation, supercooling, etc. 
Since the distance of $\mu$ from the coexistence curve is 
the important parameter, we introduce the notation 
$\Delta \mu = \mu - \mu_0$ \cite{COMMENT}. 

The phase degeneracy that leads to coexistence is lifted for 
$\Delta \mu \neq 0$. For 
$\Delta \mu > 0$ the high-density phase with $\theta$ near unity 
is the equilibrium phase, while $\Delta \mu < 0$ favors the low-density phase. 
In both cases the uniform phase which is not favored by the sign of 
$\Delta \mu$ can be extremely long-lived. The kinetics of the decay 
towards equilibrium 
of a system initially in such a long-lived {\it metastable\/} 
phase is the topic of this paper. 

The lattice-gas model does not have an intrinsic 
dynamic. However, on time scales that are long compared to 
an inverse phonon frequency, the 
approach to equilibrium can be described by a stochastic dynamic. 
Here we study {\it local\/} dynamics, in which the density is 
not conserved. Among the many possible dynamics that satisfy detailed 
balance, and thus ensure approach to equilibrium, we choose 
the single-site Glauber dynamic \cite{BIND92B} 
with updates at random sites. 
In this dynamic a proposed update of a single $c_i$ is accepted with the 
probability 
$W(c_i$$\rightarrow$$1$$-$$c_i) 
$$=$$[ 1$$+$$\exp(\Delta E / k_{\rm B} T) ]^{-1}$,
where 
$\Delta E$ is the energy change which would ensue if the update were accepted. 
Each attempted update advances the clock by one step, and we measure 
time in MC Steps per Site (MCSS).  

\section{Nucleation and Growth}
\label{sec:NG}

Next we summarize elements of the theory of 
phase transformation by nucleation and growth in systems with nonconserved 
order parameter. For further details, see
Refs.~\cite{RIKV94,RICH94,RIKV94A}. 
We concentrate on homogeneous nucleation, thus ignoring 
effects of surfaces or disorder.  Hence our 
comparisons will be with simulations using 
periodic boundary conditions and position-independent interactions and 
chemical potentials.

A central role is played by the {\it metastable 
lifetime\/}, $\langle \tau(\Delta \mu,T,L) \rangle$, 
which is the typical time it takes 
to convert half of the system volume from the metastable to the 
stable phase. For concreteness we define it as the 
mean first-passage time to $\theta = 1/2$ 
from an initial condition with all $c_i =$$1$ and $\Delta \mu < 0$. 

{}For the class of attractive lattice-gas models with 
short-range interactions studied here, the fluctuations that lead to the 
decay of the metastable phase are compact, $d$-dimensional  
droplets of radius $R$. 
The free energy of the droplet has two competing
terms: a positive interface term $\propto R^{d-1}$, and a
negative bulk term $\propto |\Delta \mu| R^d$.
The competition between these yields a critical droplet radius,
\begin{equation}
        R_c(\Delta \mu,T) \approx \frac{(d-1)\sigma(T)}
                       {|\Delta \mu| \Delta \theta_{0}(T)} \; ,
  \label{eq:Rc}
\end{equation}
where $\sigma(T)$ is the equilibrium surface tension and 
$\Delta \theta_{0}(T)$ is the equilibrium density difference at coexistence. 
Both are positive for $T<T_c$. Droplets with $R < R_c$
tend to decay, while droplets with $R > R_c$ tend to continue growing.
The free-energy cost of the critical droplet with $R = R_c$, 
relative to the uniform metastable phase, is
\begin{equation}
        \Delta F(\Delta \mu,T) = \Omega_d \sigma(T)^d
        \left( \frac{d \! - \! 1}{|\Delta \mu| \Delta \theta_{0}(T)} 
        \right)^{d-1} ,
   \label{eq:DFd-pbc}
\end{equation}
where $\Omega_d$ is a weakly $T$-dependent shape factor such that the volume 
of a droplet of radius $R$ is $\Omega_d R^d$.
Nucleation is a stochastic process, and the nucleation rate per unit volume
is given by a Van't Hoff-Arrhenius relation,
\begin{equation} 
        I(\Delta \mu,T) \propto
|\Delta \mu|^{K} \exp \left[ - \frac{\Xi(T)}
                                    {k_{\rm B}T |\Delta \mu|^{d-1}} \right]
\; ,
\label{eq:NucRate}
\end{equation}
where $\Xi(T)$ is the $\Delta \mu$-independent part of $\Delta F$.
{}For the model studied here, the prefactor exponent 
$K$ equals 3 for $d=2$ 
\cite{RIKV94,RIKV94A,LANG67,GNW80}
and is believed to be $-1/3$ for $d=3$ \cite{GNW80}. 

{}For systems of finite linear size $L$, a $T$- and $L$-dependent crossover  
called the Thermodynamic Spinodal (THSP), $|\Delta \mu|_{\rm THSP}$ 
\cite{RIKV94A} is determined by the condition $R_c \approx L$. 
Thus $|\Delta \mu|_{\rm THSP} \sim L^{-1}$. 
{}For $|\Delta \mu| < |\Delta \mu|_{\rm THSP}$, $R_c$ would exceed $L$.
This is called the Coexistence (CE) regime
because the critical fluctuation here 
resembles two coexisting domains of different density.
{}For $|\Delta \mu| > |\Delta \mu|_{\rm THSP}$ 
(but not too large, as discussed below),
the lifetime is dominated by the inverse of the total nucleation rate:
\begin{equation}
\langle \tau_{\rm SD}(\Delta \mu,T,L) 
\rangle \approx \left( L^d I(\Delta \mu,T) \right)^{-1}
\propto
L^{-d} |\Delta \mu|^{-K} 
\exp \left[ \frac{\Xi(T)}{ k_{\rm B}T |\Delta \mu|^{d-1}} \right]
\; .
\label{eq:tauSD}
\end{equation}
This is inversely proportional to the system volume, $L^d$.
The subscript SD stands for Single Droplet
and indicates that the phase transformation 
is completed by the first critical droplet which nucleates. 

The fact that a supercritical droplet grows at a finite velocity, 
$v(\Delta \mu,T)$, leads to 
a second crossover, called the Dynamic Spinodal (DSP) \cite{RIKV94A}. 
A reasonable criterion to locate the DSP is that the average time between
nucleation events, $\langle \tau_{\rm SD} \rangle$, should equal the time
it takes a droplet to grow to a size comparable to $L$. This gives the
asymptotic relation
\begin{equation}
\label{eq:HDSP}
        |\Delta \mu|_{\rm DSP}(T,L)  \sim
        \left[ \frac{\Xi(T)}
                {(d+1)  k_{\rm B}T \ln L } \right]^{\frac{1}{d-1}} .
\end{equation}
The DSP is shown in Fig.~\ref{fig:SPD} for two different values of $L$. 
The convergence to zero as $L$$\rightarrow$$\infty$ is exceedingly slow, 
so that 
the crossover is observable in sufficiently small systems 
\cite{RICH94,RIKV97B}. 
{}For $|\Delta \mu| > |\Delta \mu|_{\rm DSP}$, 
the metastable phase decays through many droplets
which nucleate and grow independently in different parts of the system.
This is called the Multidroplet (MD) regime
\cite{RIKV94A}. The KJMA theory of metastable decay in large systems 
\cite{KOLM37,JOHN39,AVRAMI}
gives the lifetime in this regime,
\begin{eqnarray}
\langle \tau_{\rm MD}(\Delta \mu,T) \rangle 
&\approx& 
\left[ \frac{\Omega_d I v^d}{(d+1) \ln 2}
            \right]^{- \frac{1}{d+1}} 
\label{eq:tauMDa} 
\\
&\propto& 
v^{{-d}\over{d+1}}
|\Delta \mu|^{{-K}\over{d+1}} 
\exp \left[ \frac{\Xi(T)}{ (d+1) k_{\rm B}T |\Delta \mu|^{d-1}} \right]
\;,
\label{eq:tauMDb}
\end{eqnarray}
which is {\em independent\/} of $L$. 
Note that the velocity depends on $T$ and $\Delta\mu$.  
Comparing Eqs.~(\ref{eq:tauSD}) and~(\ref{eq:tauMDb}), it becomes clear 
that it is useful to plot $\ln \langle \tau \rangle$ vs 
$1/|\Delta \mu|^{d-1}$. 
In such plots, which are shown in Fig.~\ref{fig:T2d3d}, 
the lifetimes in both regimes lie on approximately 
straight lines, with the 
slope in the SD regime $(d+1)$ times that in the MD regime. 

In the MD regime the order parameter 
decays according to ``Avrami's law,''
\begin{equation}
\frac{\langle \theta (t) \rangle - \theta_{\rm s}(\Delta \mu,T)}
{\theta_{\rm ms}(\Delta \mu,T) - \theta_{\rm s}(\Delta \mu,T)}
\approx 
\exp \left[ - \ln 2 
\left( \frac{t}{\langle \tau_{\rm MD} \rangle} \right)^{d+1} \right] 
\;,
\label{eq:Alaw}
\end{equation}
where $\theta_{\rm s}(\Delta \mu,T)$ and $\theta_{\rm ms}(\Delta \mu,T)$ 
are the equilibrium and quasi-equilibrium values of the density 
in the stable and metastable phases, respectively. 
Equilibrium simulations or measurements easily yield 
$\theta_{\rm s}(\Delta \mu,T)$. 
By fitting Eq.~(\ref{eq:Alaw}) to simulational or experimental data for 
$\theta (t)$, one can therefore estimate 
$\theta_{\rm ms}(\Delta \mu,T)$ and 
$\langle \tau_{\rm MD}(\Delta \mu,T) \rangle$. 
The latter gives the 
combination $I v^d$ of the nucleation rate and growth velocity 
[Eq.~(\ref{eq:tauMDa})], but not $I$ and $v$ separately. These 
quantities can, however, be resolved by using results for the two-point 
density correlation functions in the KJMA picture 
\cite{SEKI84A,OHTA87}. 
These give the 
time-dependent 
variance 
\begin{equation}
\langle \theta^2(t) \rangle - \langle \theta(t) \rangle^2 
\approx 
L^{-d} d \Omega_d \left[\langle \theta (t) \rangle - \theta_{\rm s} \right ]^2 
\Phi(I v^d t^{d+1}) \ (2 t)^d \ v^d \;,
\label{eq:Varm}
\end{equation}
where $\Phi$ is determined by numerical integration of an 
expression involving the correlation function \cite{RICH94,RAMO99}. 
The last factor 
of $v^d$ in Eq.~(\ref{eq:Varm}) is not multiplied by $I$.  
This 
permits the separate determination of $v$ 
by measurement of the variance of $\theta (t)$ 
in a series of independent simulations or experiments. 

The description of the decay in terms of droplets breaks down for 
sufficiently large $|\Delta \mu|$. The crossover 
marking the transition to this Strongly Forced (SF) regime is 
estimated as the $L$-independent ``Mean-Field Spinodal'' (MFSP), 
$|\Delta \mu|_{\rm MFSP}(T)$, for which $R_c = 1/2$. 

\section{Numerical Results}
\label{sec:NR}

\begin{figure}
\vspace*{13pt}
\centerline{
\epsfxsize=3.5in
\epsfysize=3.5in
\epsfbox{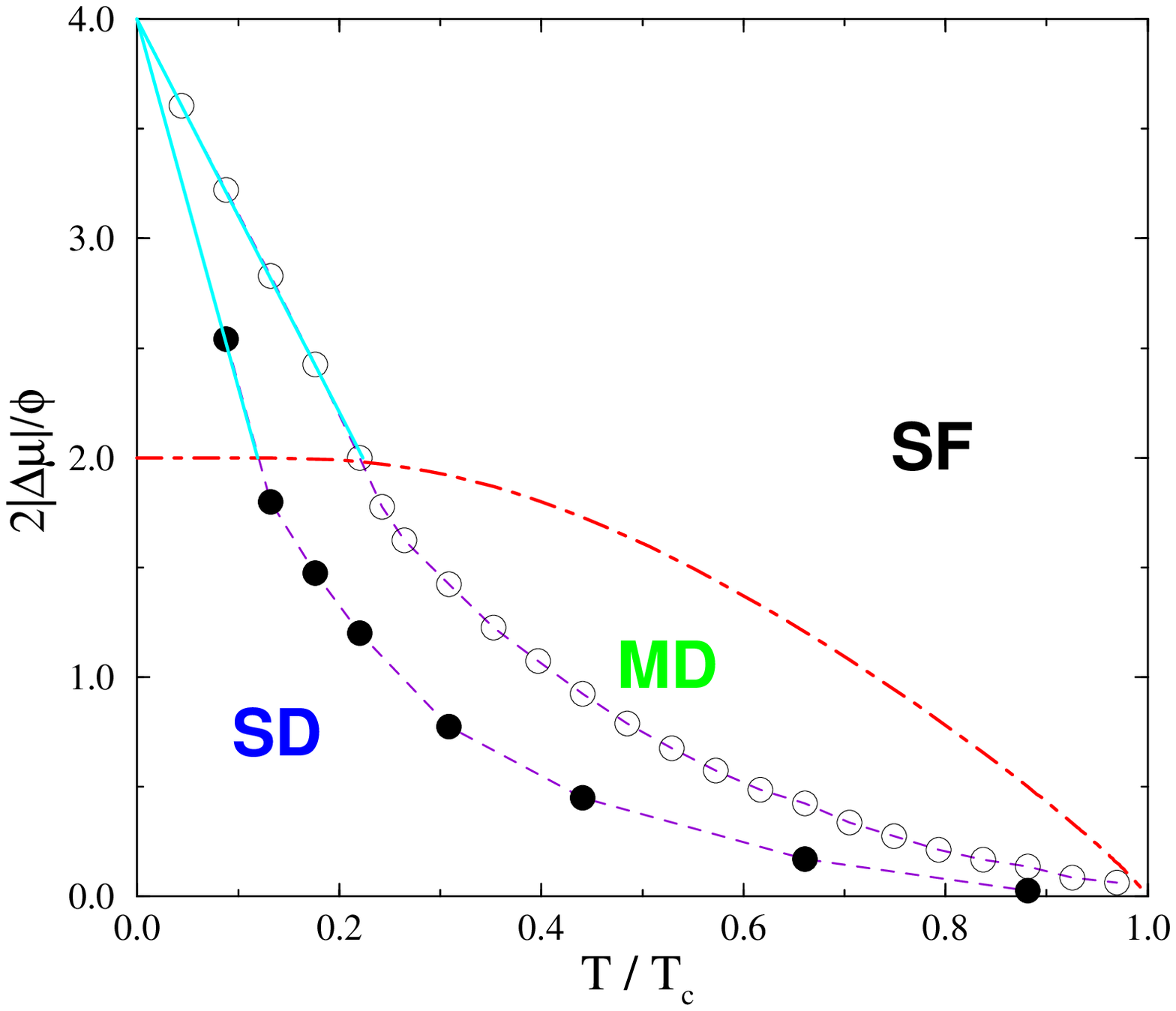}
}
\vspace*{10pt}
\caption[]{
``Metastability phase diagram'' for $d$$=$$2$, 
with the Single Droplet (SD), Multidroplet (MD), 
and Strongly Forced (SF) regimes labeled.  The 
coexistence (CE) regime for very small $|\Delta \mu|$ is not shown.  
The dot-dashed line is the estimate for the Mean-Field Spinodal (MFSP), 
as described in the text.  
The estimates for the Dynamic Spinodal (DSP) are shown for 
$L$$=$$24$ (empty circles) and $L$$=$$240$ (filled circles).  These 
data points are joined by dashed lines as guides to the eye, 
and agree with the analytical low-temperature expression 
for $2|\Delta\mu|_{\rm DSP}/\phi > 2$ 
from Eq.~(\ref{eq:LowTDSP}), marked by the solid straight lines.  
}
\label{fig:SPD}
\end{figure}

The ``metastability phase diagram'' 
for the square-lattice Ising lattice-gas model is shown in Fig.~\ref{fig:SPD}.
The three decay regimes shown are the SF, 
MD, and SD regimes.  
The CE regime, which occurs at very low $|\Delta \mu|$, is 
not shown. The MFSP, which separates the SF regime 
from the other regimes, is given by Eq.~(\ref{eq:Rc})
with exactly known expressions for both $\sigma(T)$ and 
$\Delta \theta_{0}$.  This curve goes to zero at $T_c$.  
The DSP is shown for two system sizes, $L$$=$$24$ 
and~$240$.  Our estimate for 
the location of the DSP is where the standard deviation 
of the lifetime, $\Delta\tau$, is equal to $\langle \tau \rangle/2$. 
In the SD (and CE) regimes $\Delta\tau$$\approx$$\langle \tau \rangle$, while 
$\Delta\tau \ll \langle \tau \rangle$ in the MD and SF 
regimes \cite{RIKV94A}. 

As $L$ increases, the portion of the diagram occupied by the MD regime 
increases.  
For  $2 < 2|\Delta \mu|/\phi < 4$, 
the critical droplet consists of a single lattice site 
\cite{NEVE91}.  
{}For the dynamic used here, there exists an analytical low-temperature 
estimate for the DSP in this range of $|\Delta \mu|$ \cite{JLEE94A}. 
It is obtained by equating 
the growth time and the nucleation time, and is (for $d=2$) given by 
\begin{equation}
\frac{2 |\Delta \mu|_{\rm DSP}(T,L)}{ \phi} 
=
4 - \frac{4 k_{\rm B} T}{\phi} \left( \frac{3}{2} \ln L - 0.82 \right)
\; .
\label{eq:LowTDSP}
\end{equation}
The only adjustable parameter is the last term inside the parenthesis, 
which was fit to give agreement for $L$ between $8$ and $240$.  
This expression for the DSP is marked by the two solid straight lines in 
Fig.~\ref{fig:SPD}.  
It separates the SF regime from a low-temperature 
SD regime, where the nucleating droplet occupies only a single lattice site.  
The agreement between the simulation data and the analytical 
expression is very good.  
Below the intersection with the MFSP, the low-temperature form for 
$|\Delta\mu|_{\rm DSP}$ 
merges smoothly into the form given by Eq.~(\ref{eq:HDSP}), which goes to 
zero as $T \rightarrow T_c$. 

\begin{figure}
\vspace*{13pt}
\centerline{
\epsfxsize=3.5in
\epsfysize=3.5in
\epsfbox{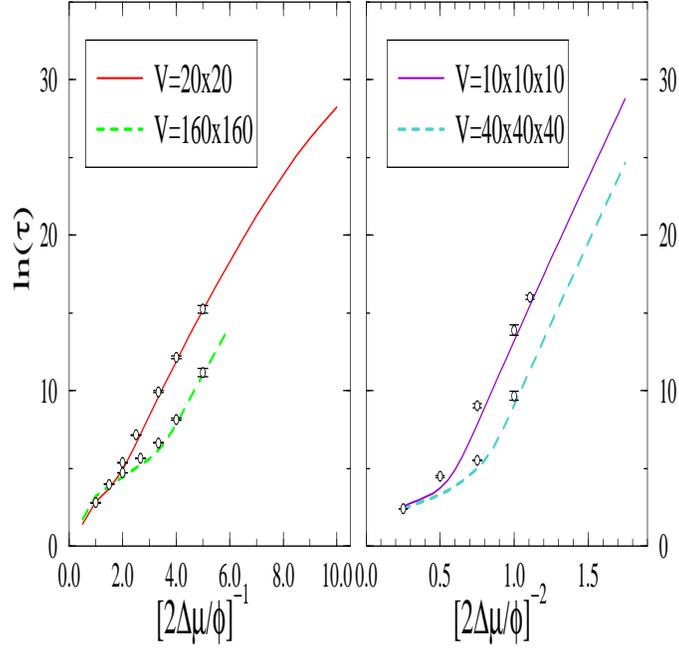}
}
\vspace*{10pt}
\caption[]{
The average lifetime of the metastable phase, measured in MCSS, shown vs 
$\left(\phi/2|\Delta\mu|\right)^{d-1}$ for different system sizes and 
dimensions. 
In both parts, the curve for the smaller system is solid and the curve 
for the larger system is dotted. 
For $d$$=$$2$ (a), the 
system sizes shown are $20^2$ and $160^2$ 
at $T$$=$$0.325$$\phi$$=$$0.57$$T_c$.  
For $d$$=$$3$ (b), the 
system sizes shown are $10^3$ and $40^3$ 
at $T$$=$$0.5$$\phi$$=$$0.44$$T_c$.  
The symbols are from actual simulations. The curves 
are PD extrapolations based on 
equilibrium data sampled for the smaller system size.  
See the text for further detail.  
The different slopes in the MD and SD regimes are clearly seen. 
After Ref.~\protect\cite{KOLE98A} with additional data. 
}
\label{fig:T2d3d}
\end{figure}

As seen from Eqs.~(\ref{eq:tauSD}) and~(\ref{eq:tauMDb}),
plotting $\langle\tau\rangle$ 
versus $1/|\Delta\mu|^{d-1}$ gives a reasonable estimate for 
$\Xi(T)/k_{\rm B}T $ in the SD regime and 
$\Xi(T)/(d+1)k_{\rm B}T $ in the MD regime.  
Examples are shown in Fig.~\ref{fig:T2d3d} for $d$$=$$2$ (a) and 
$d$$=$$3$ (b).  
Slight curvatures are present, due to prefactors in the nucleation rate 
and also to the proximity of a crossover (the THSP, DSP, or MFSP).  
Well away from crossovers, the curvatures have been used to 
estimate the prefactor exponent $K$ \cite{RIKV94,RIKV94A}.  
The symbols in both panels (with error estimates from $\Delta\tau$), 
are from 1000 simulations using the 
$n$-fold way algorithm \cite{BORT75}.  

The $n$-fold way algorithm 
is based on dividing all sites into classes reflecting the 
current state of the site and its neighbors.  It 
yields class populations 
of $n$$=$$10$ classes for $d$$=$$2$ and 
$n$$=$$14$ classes for $d$$=$$3$.  
These populations can be utilized to 
`coarse-grain' or `lump' all configurations onto the 
one-dimensional variable $\theta$.  
Measuring these class populations and 
utilizing the analytical form for the 
acceptance probabilities yields average growth and 
shrinkage rates for the stable phase.  
This is the central idea behind the 
Projective Dynamics (PD) method \cite{KOLE98A,KOLE98B,NOVO99B}, 
which was used to obtain the curves in Fig.~\ref{fig:T2d3d}.
Here we measured the class populations in equilibrium, 
which provides reasonable approximations for the actual class populations 
during metastable escape for small $|\Delta\mu|$.  
This approximation causes the slight differences between the 
curves and the measured data points.  
Using the correct measured 
class populations during the escape from the metastable phase 
gives the exact lifetime (within statistical errors) \cite{NOVO99B}.  
The PD method also enables one to obtain results for 
large systems from class populations in small systems \cite{KOLE98A,KOLE98B}.  
In fact, the PD results for the large systems in Fig.~\ref{fig:T2d3d} are 
obtained from the equilibrium class populations in the smaller systems.  
By using such advanced algorithms 
\cite{BORT75,KOLE98A,KOLE98B,NOVO99B,NOVO95}, 
one can speed up calculations of metastable lifetimes 
by many orders of magnitude --- {\it without\/} 
changing the dynamics of the system.  

The approximate KJMA theory of phase transformation kinetics is 
a remarkably versatile tool for analysis of experimental results. 
Nevertheless, 
there are very few studies in which it has been explicitly tested 
in theoretical models with microscopic dynamics on a length scale much 
smaller than $R_c$ 
\cite{RAMO99,ELDE96,SHNE98}. 
Here we present results using 
Eqs.~(\ref{eq:Alaw}) and~(\ref{eq:Varm}) to analyse the decay  
of the metastable phase in $d$$=$$2$ MC simulations 
in the MD regime 
at 0.8$T_c$ \cite{RAMO99}.  
The systems are sufficiently large to contain at least several 
hundred supercritical droplets ($L$=256 for $2|\Delta \mu|/\phi > 0.20$ and 
$L$=1024 for $0.20 > 2|\Delta \mu|/\phi > 0.12$). 

\begin{figure}
\vspace*{13pt}
\centerline{
\epsfxsize=3.5in
\epsfysize=3.5in
\epsfbox{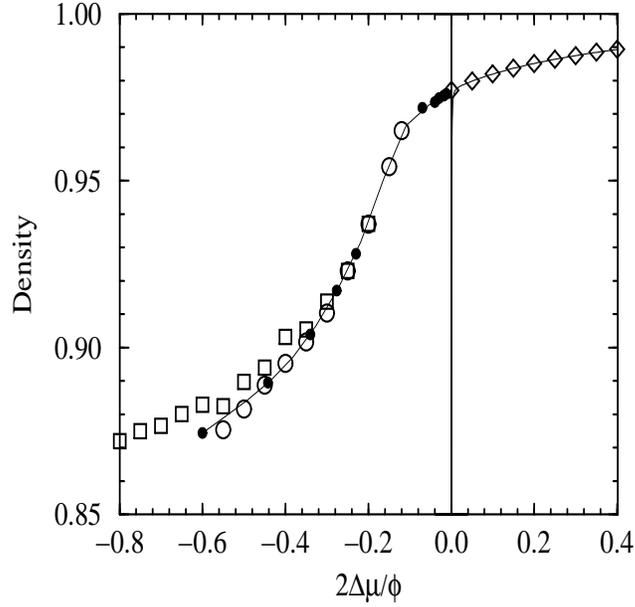}
}
\vspace*{10pt}
\caption[]{
Metastable and stable densities for $d$=2 at 0.8$T_c$.  
The equilibrium density, $\theta_{\rm s}(\Delta \mu, T)$, is shown for 
$\Delta \mu \ge 0$. The data points represented by 
empty diamonds were obtained from equilibrium 
MC simulations with $L$=256, while the thin solid curve is from 
a transfer-matrix calculation with strip width $N$=9. 
The metastable density, $\theta_{\rm ms}(\Delta \mu, T)$ is shown for 
$\Delta \mu < 0$. The data points represented by empty circles 
and squares were obtained by fitting MC data for 
$\langle \theta(t) \rangle$ to Eq.~(\ref{eq:Alaw}) as described in the text, 
while the filled circles 
represent transfer-matrix calculations with $N$ between~5 and~9. 
The thin solid curve connecting the latter points is a guide to the eye only. 
Statistical uncertainties in the MC 
results are everywhere smaller than the symbol size. 
After Ref.~\protect\cite{RAMO99}. 
}
\label{fig:M}
\end{figure}

Figure~\ref{fig:M} shows the metastable density 
$\theta_{\rm ms}(\Delta \mu,T)$ for $\Delta \mu < 0$, 
obtained by fitting data for $\langle \theta(t) \rangle$ 
to Eq.~(\ref{eq:Alaw}). 
The equilibrium density $\theta_{\rm s}(\Delta \mu,T)$ is shown for 
$\Delta \mu \ge 0$. 
As one would expect, the metastable phase becomes gradually 
less well defined as $|\Delta \mu|$ is increased. 
This is reflected by the increasing 
differences between the data points shown as empty circles and squares, 
which were obtained using different data cutoffs in the fitting procedure 
\cite{RAMO99}. For smaller $|\Delta \mu|$ these estimates coincide, 
indicating that the 
details of the fitting procedure become unimportant as the metastable phase 
becomes more well defined. 
As a check on the procedure, metastable densities measured by a 
transfer-matrix method 
\cite{RAMO99} 
are also shown. 
The agreement 
between the estimates obtained by these two unrelated methods is 
remarkable over the whole range of $|\Delta \mu|$ shown. 

\begin{figure}
\vspace*{13pt}
\centerline{
\epsfxsize=3.5in
\epsfysize=3.5in
\epsfbox{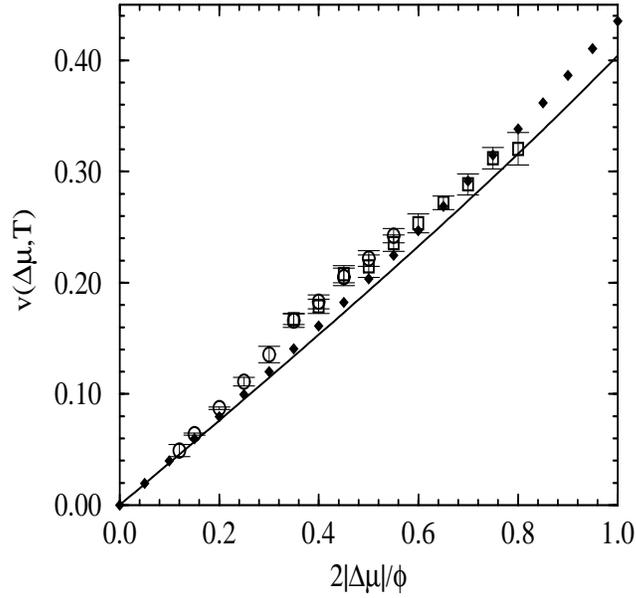}
}
\vspace*{10pt}
\caption[]{
The average propagation velocity of the interface separating the regions of 
metastable and stable phase for $d$=2 at 0.8$T_c$.  
The empty squares and circles represent estimates 
based on Eqs.~(\ref{eq:Alaw}) and ~(\ref{eq:Varm}) as described in the text. 
The error bars indicate the statistical uncertainties. 
The filled diamonds were obtained from MC simulations of a flat interface of 
length $1000$, propagating into a region of metastable phase in which 
nucleation was suppressed \protect\cite{RIKV99}.  The solid curve is 
an analytic nonlinear-response approximation \protect\cite{RAMO99,RIKV99}. 
After Ref.~\protect\cite{RAMO99} with additional data. 
}
\label{fig:V}
\end{figure}

In Fig.~\ref{fig:V} we show the average velocity of the convoluted interface 
between the metastable and stable phases, $v(\Delta \mu,T)$, obtained 
by fitting Eq.~(\ref{eq:Varm}) to the variance of the simulated 
densities, while using the values of $\theta_{\rm ms}$ and $Iv^2$ obtained 
from fitting $\langle \theta(t) \rangle$ to Eq.~(\ref{eq:Alaw}). 
As a check on 
this rather indirect way of estimating the velocity, the figure 
also contains data from a simulation of a flat 
interface propagating into a metastable bulk phase in which nucleation 
is suppressed \cite{RIKV99}, 
and a curve obtained from an analytical nonlinear-response 
approximation \cite{RAMO99,RIKV99}. The agreement between the different 
estimates is good over the whole range of $|\Delta \mu|$ shown.

\section{Discussion and Conclusion}
\label{sec:CONC}

We have shown that even for the simple case of 
homogeneous nucleation and growth applicable to the 
decay of the metastable phase in a nearest-neighbor 
Ising lattice-gas model with periodic boundary conditions, there are different 
decay regimes separated by crossovers (Fig.~\ref{fig:SPD}).  
This is because there are four length scales in the problem.  
Two length scales are fixed, the lattice spacing 
and the linear system size $L$.  
Two length scales depend on $T$ and the distance $|\Delta\mu|$ 
from the coexistence curve.  These are 
the critical droplet radius $R_c$ and the 
typical size to which a supercritical droplet grows before it 
interacts with another, 
$R_0 \approx \langle \tau_{\rm MD} \rangle v$.  
Starting from the smallest $|\Delta\mu|$, the different 
decay regimes are the CE regime where $2R_{\rm c} \ge L$, 
the SD regime where $2R_c$$\ll$$L$$\ll$$R_0$ and a single nucleating droplet 
is responsible for the decay of the metastable phase, 
the MD regime where $R_0$$\ll$$L$ and many supercritical droplets are formed 
during the decay process, and the SF 
regime where the droplet picture is not applicable.  
These regimes are separated by crossover lines, 
which depend on $T$ and $L$ in different ways.  
Between the CE and SD regimes is the Thermodynamic Spinodal with 
$|\Delta\mu|_{\rm THSP}\sim 1/L$.  
Between the SD and MD regimes is the Dynamic Spinodal with 
$|\Delta\mu|_{\rm DSP}\sim 1/(\ln L)^{1\over{d-1}}$ 
asymptotically for {\it very large\/} $L$.  
Between the MD and SF regimes is the Mean-Field Spinodal, which 
is independent of $L$.  

The average metastable lifetime $\langle\tau\rangle$ takes different 
functional forms in the different regimes (Fig.~\ref{fig:T2d3d}).  
For example, in the SD regime it is inversely 
proportional to the volume [Eq.~(\ref{eq:tauSD})], while in 
the MD regime it is independent of the volume [Eq.~(\ref{eq:tauMDb})].  
In the MD regime, ``Avrami's law'' for the time dependence 
of the order-parameter decay yields $\theta_{\rm ms}$ 
and the combination $Iv^d$ of the nucleation rate $I$ and the 
interface velocity $v$.  
However, including two-point density correlation functions 
enables one to obtain $I$ and $v$ separately.  
By fitting the KJMA theory to simulation data 
for $d$$=$$2$ we obtain results for the metastable density 
(Fig.~\ref{fig:M}) and interfacial velocity (Fig.~\ref{fig:V}) that agree 
with unrelated methods of calculating the same quantities.  
We believe the data-analysis methods discussed here should be applicable to 
many experimental systems, as well. 

\section*{Acknowledgements}
\label{sec:ACK}

We acknowledge comments on the manuscript by G.~Brown, G.~Korniss, and 
S.~J.\ Mitchell. This research was supported in part 
by National Science Foundation Grants No.~DMR-9871455 and DMR-9634873 
and by Florida State 
University through the Center for Materials Research and Technology and 
the Supercomputer Computations Research Institute 
(U.S.\ Department of Energy Contract No.~DE-FC05-85ER25000).


\begin{thebibliography}{10}
\bibitem{COMMENT}
 Equation~(\protect\ref{eq:HLG}) 
 is equivalent to the ferromagnetic Ising model,
  ${\cal H}_{\rm I} = - J \sum_{\langle i,j \rangle} s_i s_j - H \sum_i s_i$.
  Thus $T_c$ is known exactly for $d$$=$$2$ 
  and with high numerical precision for
  $d$$=$$3$. The transformations 
  are: $s_i$$=$$2 c_i$$-$$1$, $J$$=$$\phi/4$, and 
  $H$$=$$\Delta \mu /2$. For easy comparison with the Ising
  literature, $\Delta \mu$ is given in the figures in terms of 
  $2 \Delta \mu / \phi$$\equiv$$H/J$.
\end{thebibliography}

\clearpage


\end{document}